\documentclass[twocolumn]{aastex631}
\submitjournal{ApJ}

\shorttitle{Cosmic web elements from the $\beta$-skeleton}

\newcommand{\Msun}{\,{\rm M}$_{\odot}$\,}
\newcommand{\Msunh}{\,{\rm M}$_{\odot}$\,\ifmmode h^{-1}\else $h^{-1}$\fi}
\newcommand{\Mpch}{\,{\rm Mpc}\,\ifmmode h^{-1}\else $h^{-1}$\fi}
\newcommand{\kpch}{\,{\rm kpc}\,\ifmmode h^{-1}\else $h^{-1}$\fi}

\graphicspath{{./}}

\begin{document}

\title{The four cosmic tidal web elements from the $\beta$-skeleton}

\correspondingauthor{John F. Su\'arez-P\'erez, Jaime E. Forero-Romero}
\email{jf.suarez@uniandes.edu.co, je.forero@uniandes.edu.co}

\author{John F. Su\'arez-P\'erez}
\affiliation{Departamento de F\'isica, 
Universidad de los Andes,
Cra. 1 No. 18A-10, 
Bogot\'a, Colombia}

\author{Yeimy Camargo}
\affiliation{Departamento de F\'isica, 
Universidad Nacional de Colombia, 
Sede Bogot\'a, Av. Cra 30 No 45-03, 
Bogot\'a, Colombia}

\author{Xiao-Dong Li}
\affiliation{School of Physics and Astronomy, 
Sun Yat-Sen University, 
Guangzhou 510297, 
P.R. China}

\author{Jaime E. Forero-Romero}
\affiliation{Departamento de F\'isica,
Universidad de los Andes, 
Cra. 1 No. 18A-10, 
Bogot\'a, Colombia}



\begin{abstract}

Precise cosmic web classification of observed galaxies in massive spectroscopic surveys can be either highly uncertain or computationally expensive.
As an alternative, we explore a fast Machine Learning-based approach to infer the underlying dark matter tidal cosmic web environment of a galaxy distribution from its $\beta$-skeleton graph.
We develop and test our methodology using the cosmological magnetohydrodynamic simulation Illustris-TNG at $z=0$.
We explore three different tree-based machine-learning algorithms to find that a random forest classifier can best use graph-based features to classify a galaxy as belonging to a peak, filament or sheet as defined by the T-Web classification algorithm.
The best match between the galaxies and the dark matter T-Web corresponds to a density field smoothed over scales of $2$ Mpc, a threshold over the eigenvalues of the dimensionless tidal tensor of $\lambda_{\rm{th}}=0.0$
and galaxy number densities around $8\times 10^{-3}$ Mpc$^{-3}$.
This methodology results on a weighted F1 score of 0.728 and a global accuracy of 74\%.
More extensive tests that take into account lightcone effects and redshift space distortions (RSD) are left for future work.
We make one of our highest ranking random forest models available on a public repository for future reference and reuse.

\end{abstract}


\section{Introduction} \label{sec:intro}
The galaxy distribution on large spatial scales follows a structured
pattern commonly known as the cosmic web.
This web can be described as dense peaks connected by anisotropic
filaments and walls woven across vast under-dense voids \citep{Bond1996}.
The emergence of this pattern is understood as the evolution of
initial density fluctuations growing through gravitational instability
\citep{ZelDovich1970,White1987}, a picture that can be followed in
great detail with N-body cosmological simulations
\citep{Schmalzing1999,Vogelsberger2014}.

Cosmic web classification methods usually strive to classify it into four elements: peak, filament, sheet, or void \citep{Libeskind2018}.
However, most of the methods that take as an input observed galaxies focus on the detection of some, not all, cosmic web elements.
An example is the great diversity of void-finders
\citep{Aikio1998,Padilla2005,Platen2007, Neyrinck2008,Elyiv2015,Sutter2015,Xu2019} and filaments-finders
\citep{Novikov2003,Stoica2007,Zhang2009,Aragon-Calvo2010,Sousbie2010,Cautun2013,Chen2015,Luber2019,Bonnaire2020}.

The methods that classify observed galaxies into the four cosmic web elements can be roughly divided into two types according to its computational cost and global accuracy.
The first type interpolates galaxy positions on a grid to process this number density field in the same way as a dark matter density field \citep{Eardley2015,Alpaslan2016,Tojeiro2017,Shadab2019}.
This approach has a low computational cost, but it is uncertain
how accurate are their results with respect to the expected dark matter cosmic web.

A second type of algorithms first performs a full dark matter density field reconstruction on the observed galaxy distribution.
Most of these algorithms use Bayesian statistics coupled with Monte
Carlo sampling and N-body
simulations \citep{Jasche2010,Jasche2013a,Bos2014,LeclercqJasche2015,Horowitz2019,Burchett2020};
some others are based on the density distribution around halos that are associated to galaxy groups \citep{Wang2009, Munoz-Cuartas2011}.
This generic approach can provide accurate dark matter distributions but can
be very expensive from the computational point of view.

Machine Learning (ML) methods applied onto network statistics represent a middle point between these two extremes. They promise a reasonable accuracy at a low computational cost.
ML methods have already been applied to this problem of finding the four web elements
starting from the dark matter halo distribution \citep{Hui2018, Tsizh2019} and the dark matter particles \citep{2020MNRAS.497.5041B}.
For instance,
\cite{Tsizh2019} built networks by linking halos within a fixed linking length.
From these graphs, they computed ten different network metrics to feed different ML algorithms and predict the cosmic web for each halo. The true cosmic web environments were quantified using five different cosmic web classifications.

This ML approach avoids the uncertain grid interpolation of
tracers and the expensive dark matter density reconstruction.
Under this approach of supervised ML classification,
the algorithm needs a feature dataset (some halo or galaxy
properties), the labels to classify the dataset (the four cosmic web
environments) and an algorithm to link the features to the classes.

The work we present here contributes to this new line of work.
Here we explore to what extent the $\beta$-skeleton graph \citep{Fang2019} can be used to predict the T-Web \citep{Forero-Romero2009} environment classification.
We test three different supervised ML
algorithms (Classification Trees, Extra Gradient Boosting, and Random Forests)
to train from the features and predict the environment.
We perform a thorough exploration of parameter space to find the best match between the T-Web and the graph based features.
A crucial improvement of the current study is that we use galaxies as tracers.
To this end, we use the Illustris-TNG simulation \citep{Nelson2018,Naiman2018,Springel2018, Marinacci2018, Pillepich2018, Nelson2019} at redshift $z=0$.
Finally, we also release the best ML model for future reference and reuse.

This paper is organized as follows.
In Section \ref{sec:init} we describe the relevant aspects of the Illustris-TNG
simulations related to our work, the T-Web algorithm,
and the $\beta$-skeleton graph.
In Section \ref{sec:link} we describe the mechanism to link the
$\beta$-skeleton to the T-Web using machine learning algorithms.
There we explain the features and meta-parameters we used, the
classification algorithms and the metrics for model evaluation.  
In Section \ref{sec:results} we present our results. In Section \ref{sec:discussion} we explore a brief discussion comparing our results against \citet{Tsizh2019}. In Section \ref{sec:extension} we discuss about the application of our results to observational data to finally summarize our conclusions in Section \ref{sec:conclusions}.

\section{Algorithms and Simulations}\label{sec:init}

\subsection{T-Web Classification}

The tidal web (T-Web) method \citep{Hahn2007, Forero-Romero2009}
classifies the large scale structure into four cosmic web types: voids, sheets, filaments, and peaks.
This classification is based on the eigenvalues of the deformation
tensor $T_{ij}$ computed as the Hessian of the gravitational potential
\begin{equation}
    T_{ij}=\frac{\partial^2\psi}{\partial r_{i}r_{j}},
\end{equation}
where $\psi$ is a normalized gravitational potential that follows the equation
\begin{equation}
    \nabla^2 \psi = \delta,
\end{equation}
and $\delta$ is the dark matter overdensity.
This tensor has three real-valued eigenvalues. 
The cosmic web environment is defined by the number of eigenvalues
larger than a threshold value $\lambda_{th}$.
Locations with three eigenvalues larger than $\lambda_{th}$ correspond
to a peak, two to a filament, one to a sheet, and zero to a void.

Computationally speaking, to define an environment in an N-body
simulation we implement the following seven steps: 1) interpolate the
mass particles with a Cloud-In-Cell (CIC) scheme over a grid to
obtain the density, 2) smooth it with an isotropic Gaussian filter,
3) compute the overdensity, 4) find the normalized potential with Fast
Fourier Transform methods, 5) compute the deformation tensor using
finite differences, 6) find the eigenvalues, and finally 7) count the
number of eigenvalues larger than the threshold $\lambda_{th}$.

\begin{figure*}
\centering
 \includegraphics[width=\textwidth]{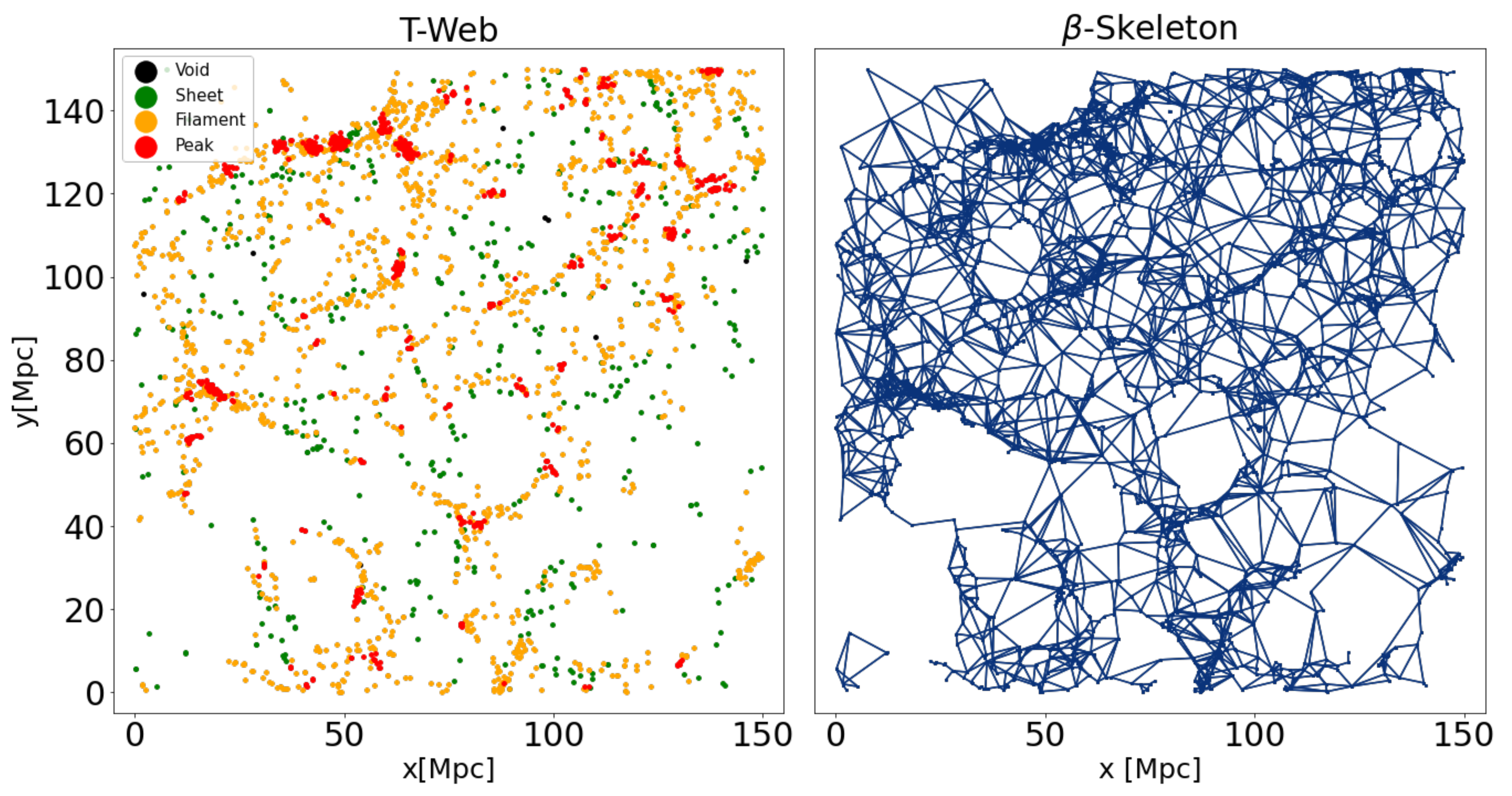}
 \caption{Comparison between the galaxies classified by its T-Web environment
   (left) and their $\beta$-skeleton computed with $\beta$=1 (right).}
 \label{fig:TWebBsk}
\end{figure*}

\subsection{The $\beta$-skeleton algorithm}

The $\beta$-skeleton is an algorithm that computes a graph over a distribution of nodes.
This algorithm depends on a positive and continuous parameter $\beta$
that defines an exclusion region between two nodes.
Here we use the Lune-based definition for the exclusion region \citep{Kirkpatrick1985}.

For a node-set $\mathbf{S}$ in a 3-dimensional Euclidean space,
the $\beta$-skeleton defines an edge set so that any two points $p$ and $q$ in $\mathbf{S}$ are connected if there is not a third point into an exclusion region. This region is defined as the intersection between two congruent spheres with diameter $\beta d$, centered at p+$\beta$(q-p)/2 and q+$\beta$(p-q)/2.
The centers are located on the axis joining the nodes $p$ and $q$.

For  $\beta=1$, the exclusion region is a sphere with a diameter $d$ with the two points under consideration located opposed to each other on the surface of that sphere.
This case receives the name of Gabriel Graph, while the $\beta$-skeleton with $\beta=2$ is known as the Relative Neighbor Graph (RNG).
The results in this paper are based on the Gabriel Graph.

More details on the properties of the $\beta$-skeleton applied to large-scale structure data can be found in \cite{Fang2019} and \cite{Garcia-Alvarado2020}.

\subsection{Illustris-TNG}

We test our algorithms on a simulation from the Illustris-TNG project. This project is a series of large,
cosmological gravo-magnetohydrodynamical simulations that follows the coupled evolution of dark matter, gas, stars, and black holes from redshift $z=127$ to $z=0$ \citep{Nelson2018,Naiman2018,Springel2018, Marinacci2018, Pillepich2018, Nelson2019}.

These simulations are based on the cosmological standard model
$\Lambda$ Cold Dark Matter (CDM) with the following cosmological parameters \citep{Ade2016}: cosmological constant
$\Omega_{\Lambda,0}$=0.6911, matter density $\Omega_{m,0}$=0.3089,
baryonic density $\Omega_{b,0}$=0.0486, normalization
$\sigma_8$=0.8159, spectral index $n_s$=0.9667 and Hubble constant
$H_o$= 100$h$ km s$^{-1}$ Mpc$^{-1}$ with $h$=0.6774.
These simulations used the \texttt{AREPO} moving-mesh code \citep{Springel2011}.
The galaxy formation physics included prescriptions for star formation and associated supernovae, magnetic fields, stellar evolution and feedback with galactic out-flows, chemical enrichment, primordial and metal-line radiative cooling of the gas, and black hole accretion and feedback \citep{Weinberger2017, Pillepich2018a}.

The Illustris-TNG suite includes simulations with boxes of sizes $50$ Mpc, $100$ Mpc, and $300$ Mpc at different resolutions.
In this work, we use the largest simulation volume with the highest mass resolution, named Illustris TNG300-1, at $z=0$.
It consists of a cube of $302.6^3$ Mpc$^3$ in volume, with a baryonic mass resolution of $8.8 \times 10^{7}$ \Msun , a dark matter resolution of $4.7 \times  10^{8}$ \Msun , and a $2500^3$ dark matter particles \citep{Nelson2019}.
We use the galaxy catalogs from the public release \citep{Pillepich2018a} to build the $\beta$-skeleton graph.
Moreover, we use the dark matter particles at $z=0$ to obtain the dark matter T-Web classification.
More extensive tests that take into account lightcone effects and redshift space distortions (RSD) are left for future work.

\begin{figure*}
    \includegraphics[width=\textwidth]{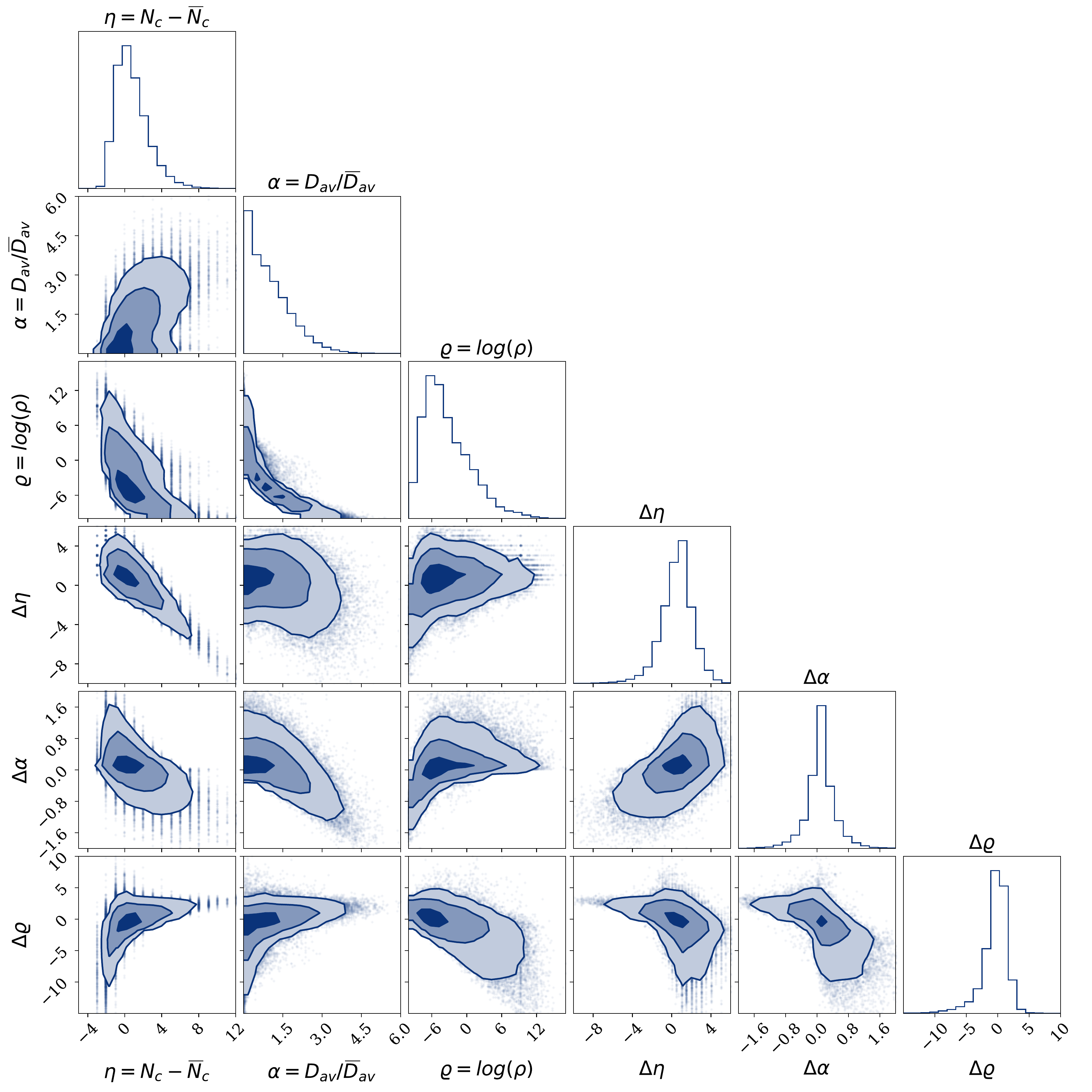}
    \caption{Histograms and correlations curves for the six features used to train the classifiers. 
	$\eta$ is the number of connections minus the global median number of connections, $\alpha$ is the ratio between the average edge normalized by the global average edge length, $\varrho$ is the logarithm of the pseudo density $\rho$, the inverse of the volume of the ellipsoid that best fit the distribution of first neighbors. 
    $\Delta\eta$, $\Delta\alpha$ and $\Delta\varrho$ are the differences over the first neighbours of $\eta$, $\alpha$ and $\varrho$, respectively. 
    The contour lines in the 2D histograms represent the levels that include 39\%, 86\% and 98\% of the data corresponding to the $1$-$\sigma$, $2$-$\sigma$ and $3$-$\sigma$ reference contour values for a 2D Gaussian probability density.}
    \label{fig:features}
\end{figure*}

\section{Linking the $\beta$-skeleton to the T-Web}\label{sec:link}

Figure \ref{fig:TWebBsk} illustrates the correlation between the T-Web classification of galaxies and their corresponding $\beta$-skeleton for a sub-box of the TNG simulation with a volume of $150$Mpc$\times 150$Mpc$\times 20$Mpc.
This spatial correlation suggests that graph derived features might have
a good chance to predict the dark matter web environment.
However, there is a variety of free parameters combinations that have to be explored to find the best match between the $\beta$-skeleton and the T-Web.
For instance, the T-Web classification depends on the smoothing scale, $\sigma$, and the threshold parameter, $\lambda_{th}$; the $\beta$-skeleton changes with $\beta$; and the number density of tracers we use to build the $\beta$-skeleton can also change.
In what follows we explore those aspects.

\subsection{T-Web and $\beta$-skeleton free parameters}

We perform the T-Web computation over a cubic grid with $256^3$ cells.
This corresponds to a cell size around of $1.18$ Mpc.
We explore five different values for the smoothing parameter $\sigma =
0.5$, $1.0$, $1.5$, $2.0$ and $2.5$, expressed in units of the cell size.
For each smoothing length we take six different values for
the eigenvalue threshold
$\lambda_{th}=0.0, 0.1, 0.2, 0.3, 0.4$ and $0.5$.

For the galaxy selection function we use a simple stellar mass cut ($\log_{10} (M^{*} / M_{\odot}\ h^{-1})>M^{*}_{\rm{lim}}$) with $M^{*}_{\rm{lim}}$: $8$, $8.5$, $9$ and $9.5$.
These thresholds in stellar mass  correspond to $515334$, $349485$, $221279$ and $129756$ galaxies, which translate into number densities of $20\times10^{-3}$, $13\times 10^{-3}$, $8\times 10^{-3}$ and $4\times 10^{-3}$ Mpc$^{-3}$, respectively.

We find that the $\beta$ skeleton features for $\beta=1$ are sufficient to make the T-Web environment prediction, i.e. adding features computed for different values of $\beta$ do not significantly improve the results.
If we include the features for $\beta=2$, the weighted F1 score of our best model only increases by 0.003.
Including features for $\beta=3$ the increase of F1 is 0.001.
Therefore, we keep $\beta=1$ fixed for the results reported in this paper.
This choice is supported by a recent study that showed that the $\beta$-skeleton entropy has its largest value for $\beta=1$ \citep{Garcia-Alvarado2020}, meaning that the connectivity of this graph is larger than other $\beta$-skeletons with $\beta>1$.
For more details on the meaning of the $\beta$-skeleton reader we refer the reader to \cite{Garcia-Alvarado2020}.

Once the four free parameters $\beta$,
$\sigma$, $\lambda_{th}$ and $M^{*}_{\rm{lim}}$ are chosen the $\beta$-skeleton and the T-Web are fixed.
Next, we define the input features to predict the four cosmic web elements.

\subsection{Galaxy Features}
We use six features for each galaxy to train the ML algorithm:
\begin{enumerate}
\item[1)]
The number of connections minus the global median number of
connections, $\eta = N_c - \bar{N}_c$.
\item[2)]
The ratio between the average edge length to first neighbors
normalized by the global average edge length, $\alpha=D_{av}/\bar{D}_{av}$.
\item[3)]
The logarithm of the pseudo-density over the graph,
  $\varrho=\log_{10}\rho$.
Where the pseudo-density $\rho$ is defined as the inverse of the volume
computed from the ellipsoid that best fits the distribution of first
neighbors over the graph.
If a galaxy has less than three neighbors its pseudo-density is fixed to be the $D_{av}^{-3}$.
\end{enumerate}
\noindent
We define three more features that we call $\Delta$-features.
Their values are computed as the difference between the corresponding
value in a node and the average value over first neighbor nodes.
They can be loosely interpreted as a divergence computed over the graph.

\begin{enumerate}
\item[4)] $\Delta$ over the number of connections, $\Delta\eta$.
\item[5)] $\Delta$ over the normalized edge length, $\Delta\alpha$.
\item[6)] $\Delta$ over the the pseudo-density, $\Delta\varrho$.
\end{enumerate}

Figure \ref{fig:features} shows the correlations among all features.
The diagonal panels present the distribution of a given parameter.

Some of these $\beta$-skeleton topological features were studied in previous works. For instance, \cite{Fang2019} used the average connection length to characterize the size and anisotropy of the Large Scale Structure.

Also, when we include the stellar mass $M^{*}$ as a feature, the weighted F1 score only increases by 0.001.
Therefore, we do not include the stellar mass as a feature.

\subsection{Classification Algorithms}

We arranged experiments with three supervised classification algorithms: Random Forests (RF), Extra Gradient Boosting (XGB) and Classification Trees (CT).
We use Scikit-Learn \citep{Pedregosa2011} Python's implementation of RF and CT and Xgboost \citep{Chen2016} Python's implementation for XGB.

Details on the implementation of these machine learning models can be found in \cite{Nordhausen2009}.
For CT we change the maximum tree depth $MD$ from 1 to 30.
For XGB we use a tree depth $MD$ from 1 to 30.
For RF we use a fixed maximum tree depth of 10 and change the number of estimators $NE$ (Trees) from 1 to 100.

\subsection{Model evaluation}

We partition the simulation box into eight cubic sub-boxes of equal volume.
For each sub-box we compute the $\beta$-skeleton and the related feature parameters.
Afterwards, we discard the galaxies located within 5 Mpc away from the sub-box limits to avoid boundary effects coming from incomplete $\beta$-skeleton information.
The T-Web information comes from the computation over the full box to take advantage of the periodic boundary conditions.

As the \emph{training} dataset we use four sub-boxes that only touch on one vertex.
The \emph{validation} dataset is composed by the features from two more sub-boxes, this dataset is used to select what T-Web parameters and algorithm meta-parameters provide the best results.
Once the free parameters and meta-parameters are fixed we use one sub-box as \emph{test} dataset to report the final scores and confusion matrix.
The remaining sub-box is not used in this paper, but is made available in a public repository\footnote{\url{https://github.com/jsuarez314/cosmicweb_bsk}} together with the best ML model.

As performance metric to compare our models we use the F1 score which is the harmonic mean between the purity (also called precision) and completeness (also called recall).
We compute the F1 score separately for each web environment.
Given the unbalance across classes, we use a weighted F1 average to report a global F1 metric.
We also use a global accuracy (the fraction of total correct predictions) in order to perform a comparison against the results by \cite{Tsizh2019}.

\begin{figure}
    \includegraphics[scale=0.55]{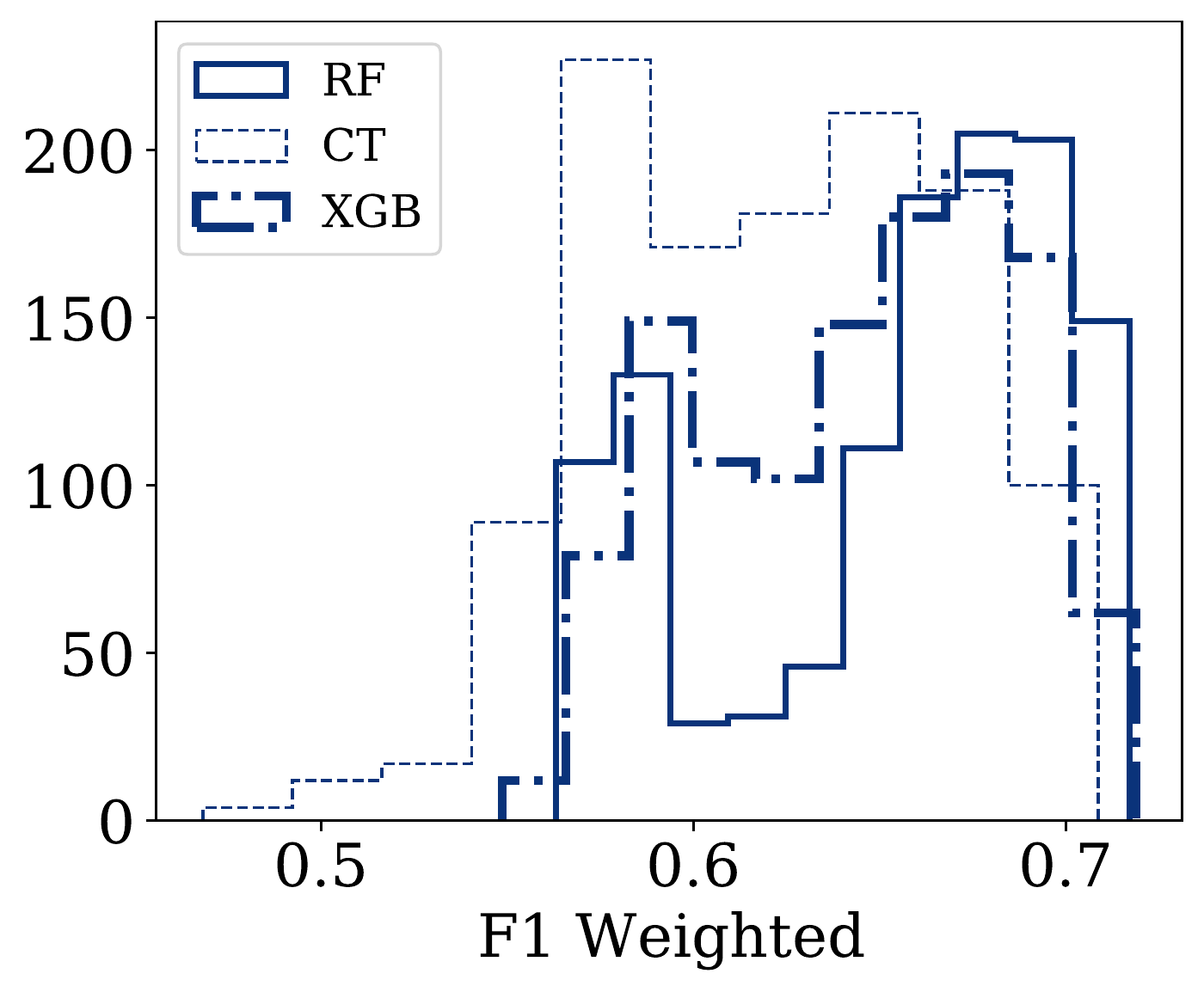}
    \caption{Histograms of the weighted F1 score for the three machine
      learning algorithms (Random Forest (RF), Extra Gradient Boosting (XBG) and Classification Trees (CT)).
      Each F1 value correspond to a different set of free parameters and metaparameters.}
    \label{fig:methods}
\end{figure}

\begin{figure*}
\centering
    \includegraphics[scale=0.24]{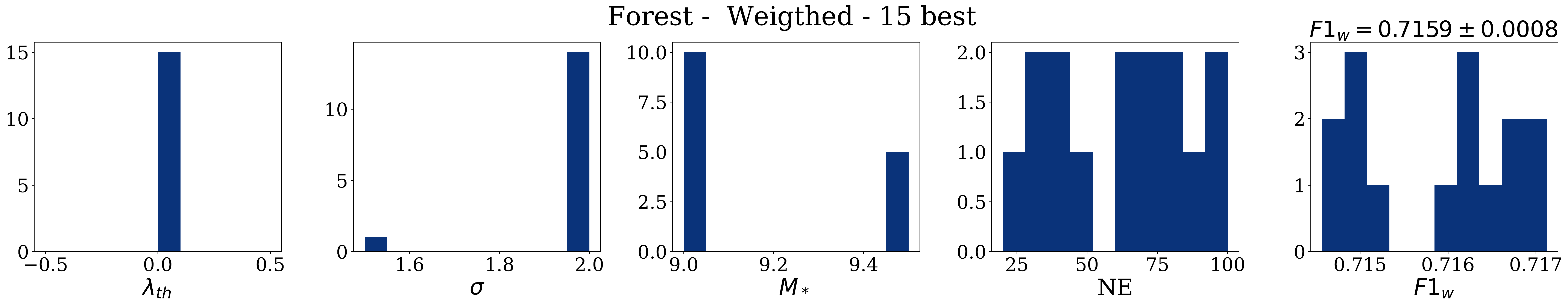}
    \caption{Parameter and meta-parameter distributions for the 15 RF models with the highest weighted F1 score.
    From these results we choose $\lambda_{th}=0.0$, $\sigma = 2.0$ and $M^{*}_{\rm lim} = 9$ as the best parameters to match the $\beta$-skeleton and the T-Web. $NE$ is the number of estimators (Trees) used in the RF algorithm.} 
    \label{fig:features_score}
\end{figure*}

\begin{table}
\centering
\begin{tabular}{cccc}
\hline
 F1-score         & Classification & Random  & Extra Gradient \\
 & Trees & Forest & Boosting \\
\hline
 $F1_{peaks}$     & 0.55 $\pm$ 0.24 & 0.55 $\pm$ 0.28 & 0.57 $\pm$ 0.24 \\
 $F1_{filaments}$ & 0.69 $\pm$ 0.07 & 0.72 $\pm$ 0.05 & 0.71 $\pm$ 0.05 \\
 $F1_{sheets}$    & 0.50 $\pm$ 0.12  & 0.53 $\pm$ 0.13 & 0.53 $\pm$ 0.11 \\
 $F1_{voids}$     & 0.22 $\pm$ 0.18 & 0.26 $\pm$ 0.19 & 0.27 $\pm$ 0.17 \\
 $F1_{weighted}$ & 0.62 $\pm$ 0.05 & 0.65 $\pm$ 0.04 & 0.64 $\pm$ 0.04 \\
\hline
\end{tabular}
\caption{Mean values and standard deviations
  over the weighted F1 scores for all models with different meta-parameters.
  The first four rows differentiate the results for each environment.
  The fifth row corresponds to the weighted F1 score over all environments.}
\label{table:elements}
\end{table}

\section{Results}\label{sec:results}

\subsection{ML algorithm performance}

We start by comparing the performance of the different ML algorithms when the T-Web and the $\beta$-skeleton parameters change.
Figure \ref{fig:methods} shows the weighted F1 distribution for
all the classification experiments.
The histograms are discriminated into the three ML algorithms.
From these three models, the results from the RF are better; the F1 value distribution is skewed towards higher values than the distribution from the CT and XGB.
From this overall response over all possible models, RF emerges as the best ML algorithm.
We can confirm this result by performing a more detailed analysis by
taking a look at the results for each cosmic web element separately.

Table \ref{table:elements} summarizes the F1 performance split by cosmic
web element in the first four rows.
The mean values and standard deviations are computed over all the ML experiments.
From this table, we confirm that the RF classifier performs
consistently better than CT for every web element.
For three out of the four web elements, the RF produces better weighted F1 scores by 0.03 that CT. The results for XGB and RF are comparable but the F1 weighted is better in the RF case.

The improved performance by the RF over the other algorithms is an expected generic outcome when the number of instances is larger than the number of features, as it is our case \citep{ali2012random,Neira2020}. The RF algorithm uses multiple CT making possible the description of more complex boundaries in feature space to separate classes, while the use of a random set of features makes it robust against over-fitting.

\subsection{Performance across web elements}

Table \ref{table:elements} also shows that some cosmic web elements
are easier to predict than others.
Focusing our attention on the RF classifier, the algorithm with the
best results, we readily observe that filaments is the environment
with the best mean F1 score (0.72).
This score is followed by peaks and sheets (0.55 and 0.53,
respectively) and finally by voids (0.26).

Void is the hardest class to predict.
The reason is the class unbalance.
There are too few galaxies in voids.
Across all models, less than $0.3\%$ of all galaxies
are classified as void galaxies.
This has two consequences.
First, the low number of instances makes it hard for the ML algorithm to
define the region in the parameter space occupied by voids.
Second, a small number of void galaxies that are misclassified is translated into large changes in the F1 score.

In comparison, Filament is the class with the best score.
Its F1 score is the highest across all four
web elements with an average value of 0.72.
Notably, the F1 dispersion across models is also the lowest corresponding to 0.05.
The reason also is the class unbalance.
Galaxies in filaments are the most represented class,
approximately $54\%$ of all galaxies are classified as filament galaxies.
This helps to explain that it is easier for the ML algorithm to learn
the characteristics that make a galaxy sit in a filament.
Peak and Sheet have intermediate F1 scores between
those for void and filament, with average values of 0.55 and 0.53.
The fraction of galaxies in peaks and sheets are also intermediate around $31\%$ and $15\%$, respectively.

\subsection{Best match between T-Web and $\beta$-skeleton}

We now focus on understanding what parameters produce the best match between the T-Web and the features derived from the $\beta$-skeleton.
In this exploration, we only use the RF algorithm because it provides the best classification results.
To do that we choose the best $N=15$ models in terms of their weighted F1 across classes and see what do they have in common.
Choosing $N=10$ or $N=20$ only changes the weighted F1 score by 0.001.

Figure \ref{fig:features_score} summarizes the parameter and meta-parameter
distribution for the best $N=15$ RF models.
From these 15 best models we pick as the best T-Web parameters $\lambda_{th}=0.0$ and $\sigma=2.0$.
These values are close to the expected range for a good visual cosmic web classification \citep{Hahn2007,Forero-Romero2009,Bustamante2015}.
The cosmic web produced by those parameters turns out to be well matched by the $\beta$-skeleton produced by galaxies with an stellar mass cut of $M^{*}_{\rm lim}=9$.
Galaxy populations with different number densities, i.e. a different threshold for the stellar mass, produce a worse match between the T-Web and the $\beta$-skeleton properties.

Finally, we select $NE=80$ the meta-parameter in the RF algorithm that gives the highest F1 score after all the other parameters have been fixed.

\subsection{Confusion Matrix and Feature Importance}

\begin{figure}
\centering
    \includegraphics[scale=0.35]{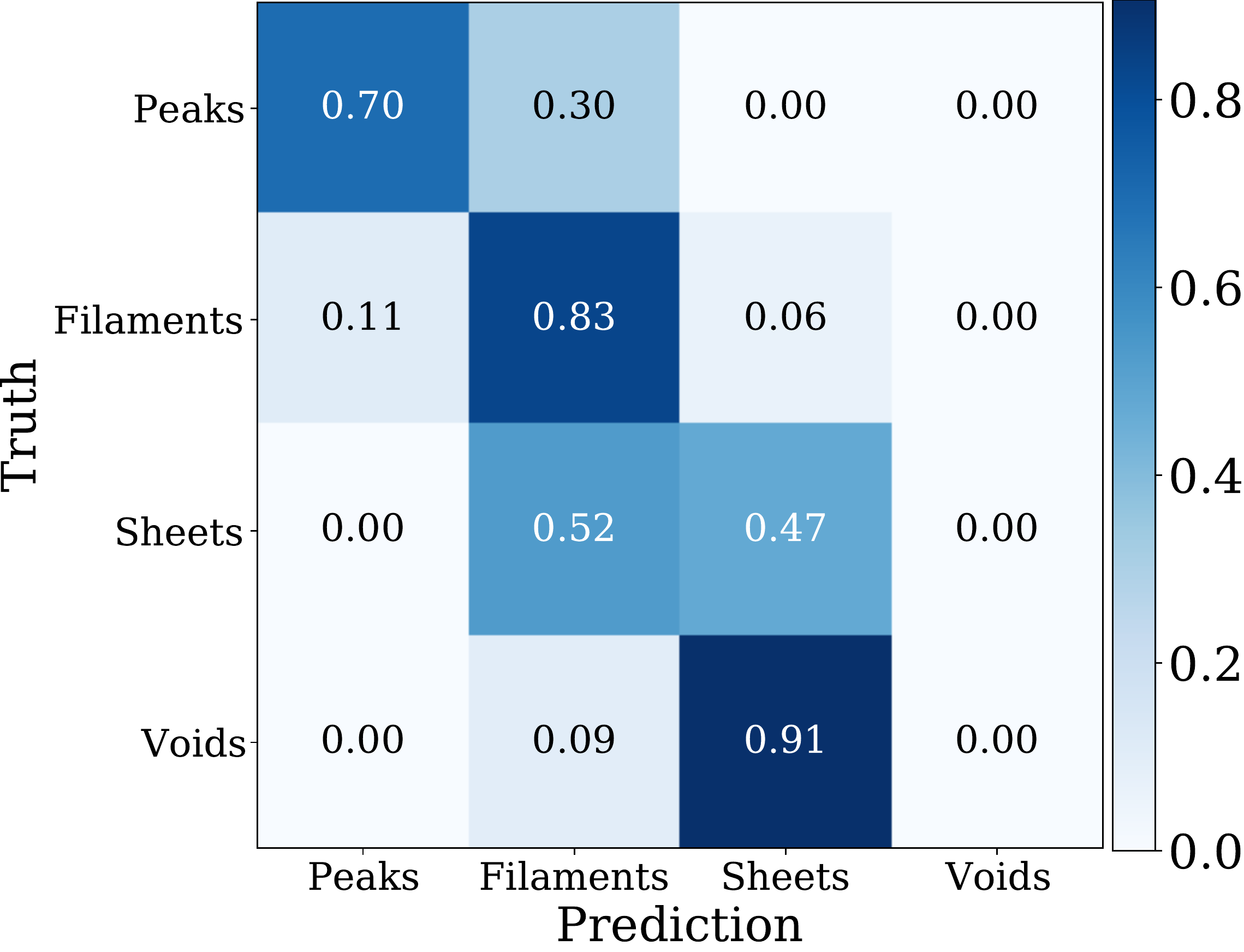}
\caption{Confusion matrix for the model
      \{$\lambda_{th}=0.0$, $\sigma=2.0$, $M^{*}_{\rm lim}=9$, $NE=80$\} computed on the \emph{test} dataset.}
      \label{fig:confusion_matrix}
\end{figure}

Figure \ref{fig:confusion_matrix} shows the confusion matrix computed on the \emph{test} dataset using our preferred model.
The numbers are the fraction of objects that correspond to
the \verb"Truth" class but are classified into the \verb"Prediction" class.
What are the most common misclassifications?
In first place comes the $91\%$ of void galaxies that are misclassified as sheet galaxies.
In second place we have the $52\%$ of sheets that are misclassified
as filaments, and finally, the $30\%$ of peaks that are misclassified as filaments.
Less than $0.01$\% of all galaxies end up classified as void galaxies by the RF algorithm.
For this model the weighted F1 score is 0.728 and its global accuracy is $74\%$.

\begin{figure}
    \includegraphics[scale=0.29]{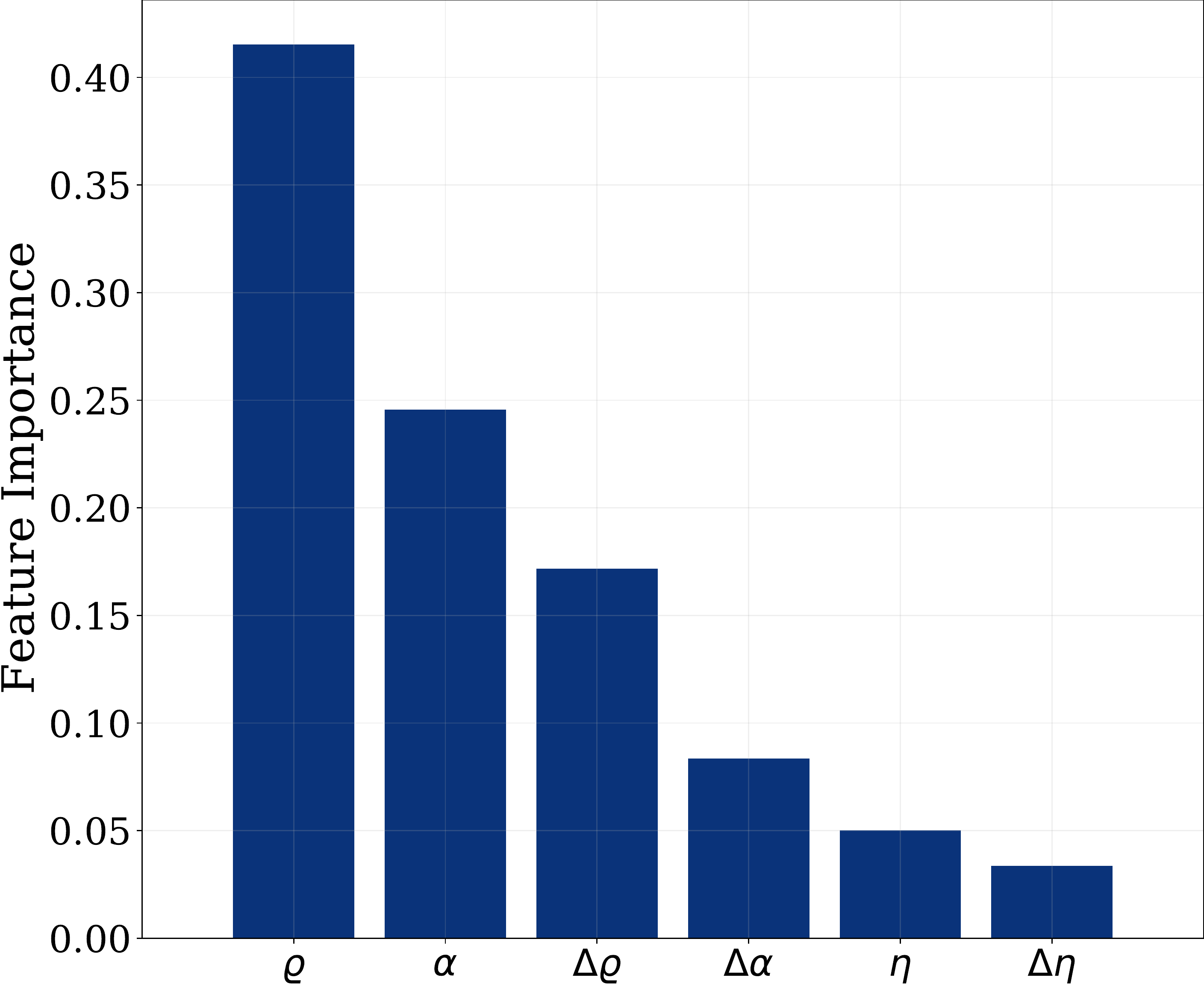}
    \caption{
      Feature importance in the RF classifier for the model
      \{$\lambda_{th}=0.0$, $\sigma=2.0$, $M^{*}_{\rm lim}=9$, $NE$=80\} computed on the \emph{test} dataset.}
    \label{fig:feature_importance}
\end{figure}

\begin{figure*}
  \centering 
    \includegraphics[width=\textwidth]{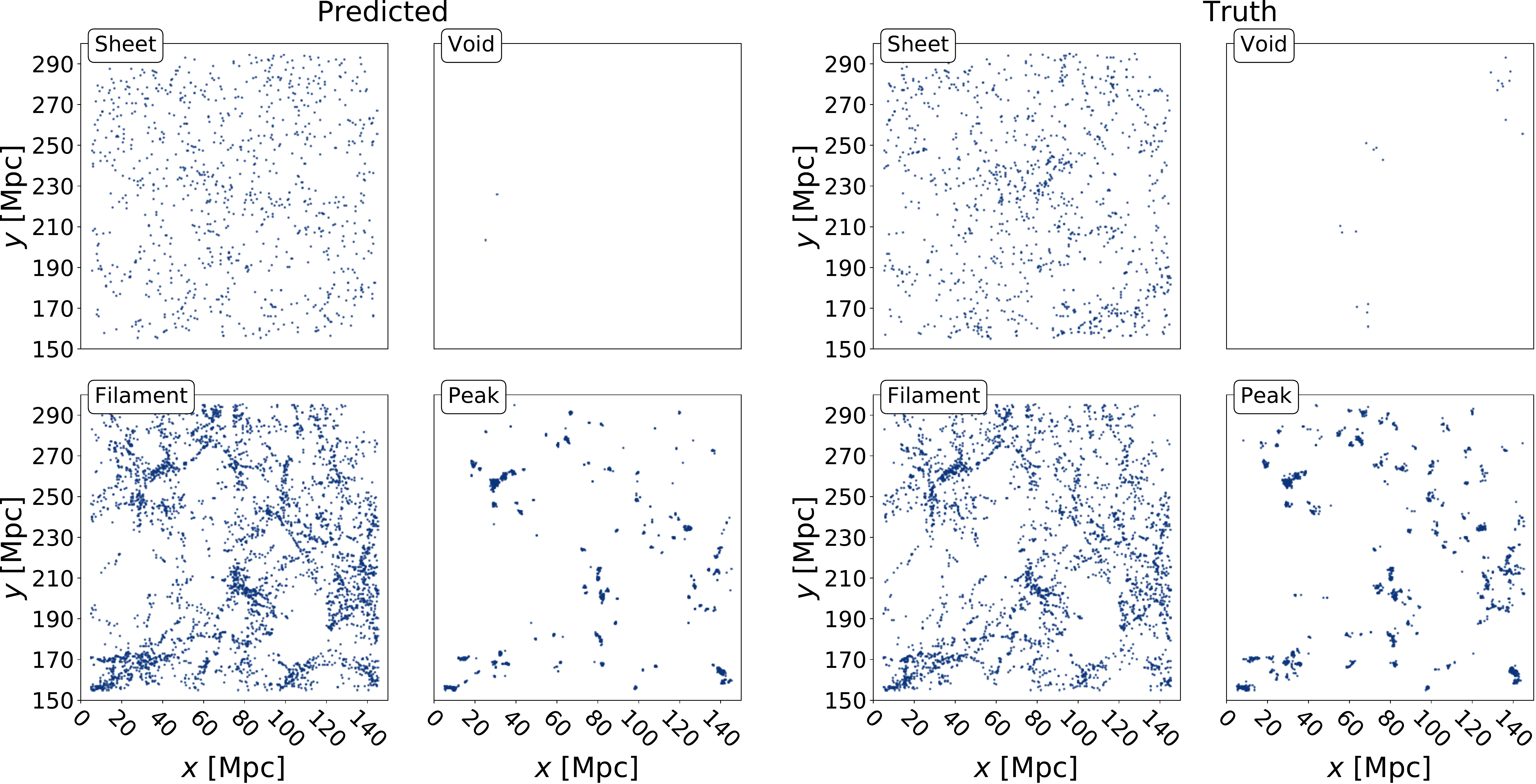}
    \caption{Spatial distribution for the galaxies classified into the
      four web elements as predicted from the best model (left) and its corresponding truth (right).}
    \label{fig:prediction}
\end{figure*}

The most common missclassifications highlights the difficulty to find void galaxies.
This turns out to be an almost impossible task for one simple reason already mentioned before:
there are very few void galaxies.
Only $0.3\%$ of the galaxies in the sample are in voids.
The difficulty to train ML algorithms to find instances under-represented classes is a generic result not only in the cosmic web context \citep{Tsizh2019} but also in other astronomical problems such as transient classification \citep{Bloom2012, Neira2020}

The second kind of misclassification shows that the galaxies classified as belonging to a filament, actually belong to neighboring environments: sheets and peaks.

These misclassifications can also be understood in terms of the topological properties of the cosmic web elements.
Voids are surrounded by sheets (hence the misclassification between the two), sheets have filaments inside them and
filaments in turn have peaks inside them \citep{Cautun2014} (explaining the misclassification of peaks and sheets into filaments).
In this progression the average density increases and the tidal field
anisotropy goes from being isotropic (voids) to anisotropic (sheets
and filaments) and isotropic (peaks) \citep{Bustamante2015}.

The ranking of correct classifications (the diagonal over the confusion matrix, which also correspond to the completeness) naturally follows the F1 trends shown in Table \ref{table:elements}.
Filaments are the class most correctly classified ($83\%$) followed by
peaks ($70\%$) and sheets ($47\%$).
Void galaxies have the worst results, with zero correct classifications.
This ranking also follows the trend in the fraction of instances belonging to
filaments ($54\%$), peaks ($31\%$), sheets ($15\%$), and voids ($0.3\%$).
A higher number of instances translates into a better chance for the
algorithm to give a correct classification.

Figure \ref{fig:feature_importance} shows the feature
importance for the classification.
As it could probably be expected, the pseudo-density ($\varrho$) and the average connection length ($\alpha$) turn out to be crucial for the classification.
The other feature with importance near to the average connection length is $\Delta \varrho$, the local change in the average pseudo-density.

Figure \ref{fig:prediction} summarizes our results in a qualitative fashion.
It shows the spatial distribution for the galaxies into the four web elements.
The left panel correspond to the predictions from the best model, the right panel correspond to the truth.
The visual impression for filament galaxies is the most
similar between truth and prediction, as expected by the quantitative results presented here.
Sheet galaxies in the prediction appear less clustered
than they should.
The failure in the void classification is clearly visible as galaxy deficit in the prediction.

\subsection{Tests to compensate for class unbalance in voids}

\begin{figure}
\centering
    \includegraphics[scale=0.35]{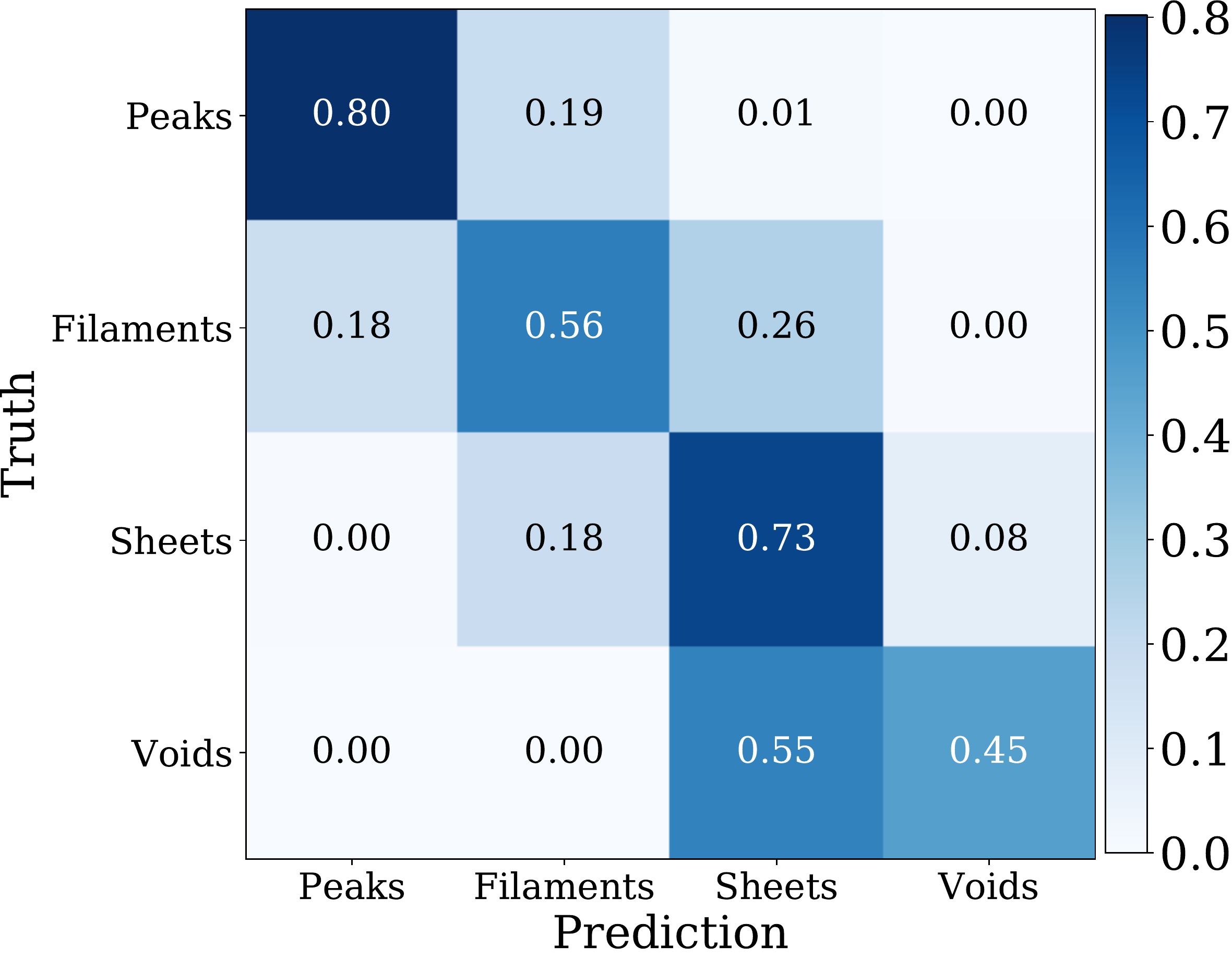}
\caption{Confusion matrix for the best Random Forest Model
      \{$\lambda_{th}=0.0$, $\sigma=2.0$, $M^{*}_{\rm lim}=9$, $NE=80$ \} using algorithmic up-weighting to compensate for the class imbalance.} 
      \label{fig:confusion_matrix_cw}
\end{figure}

We explore two strategies to improve the results for void classification.
In a first strategy we change the loss function to up-weight these galaxies.
To do this,  we set the \texttt{class\_weight} parameter to `balanced' for the parameters in the best RF model. 
With this change the model weights the classes according to the abundance in each class. 
Figure \ref{fig:confusion_matrix_cw} shows the confusion matrix for this strategy. 
With this modification the correct classification for peaks increases from $70\%$ to $80\%$, for sheets from $47\%$ to $73\%$ and for voids from $0\%$ to $45\%$. 
However, for filaments the correct classification drops from $83\%$ to $56\%$. 
Overall, the weighted F1 score drops from $0.73$ to $0.62$.
For this reason we discard this open.

\begin{figure}
\centering
    \includegraphics[scale=0.35]{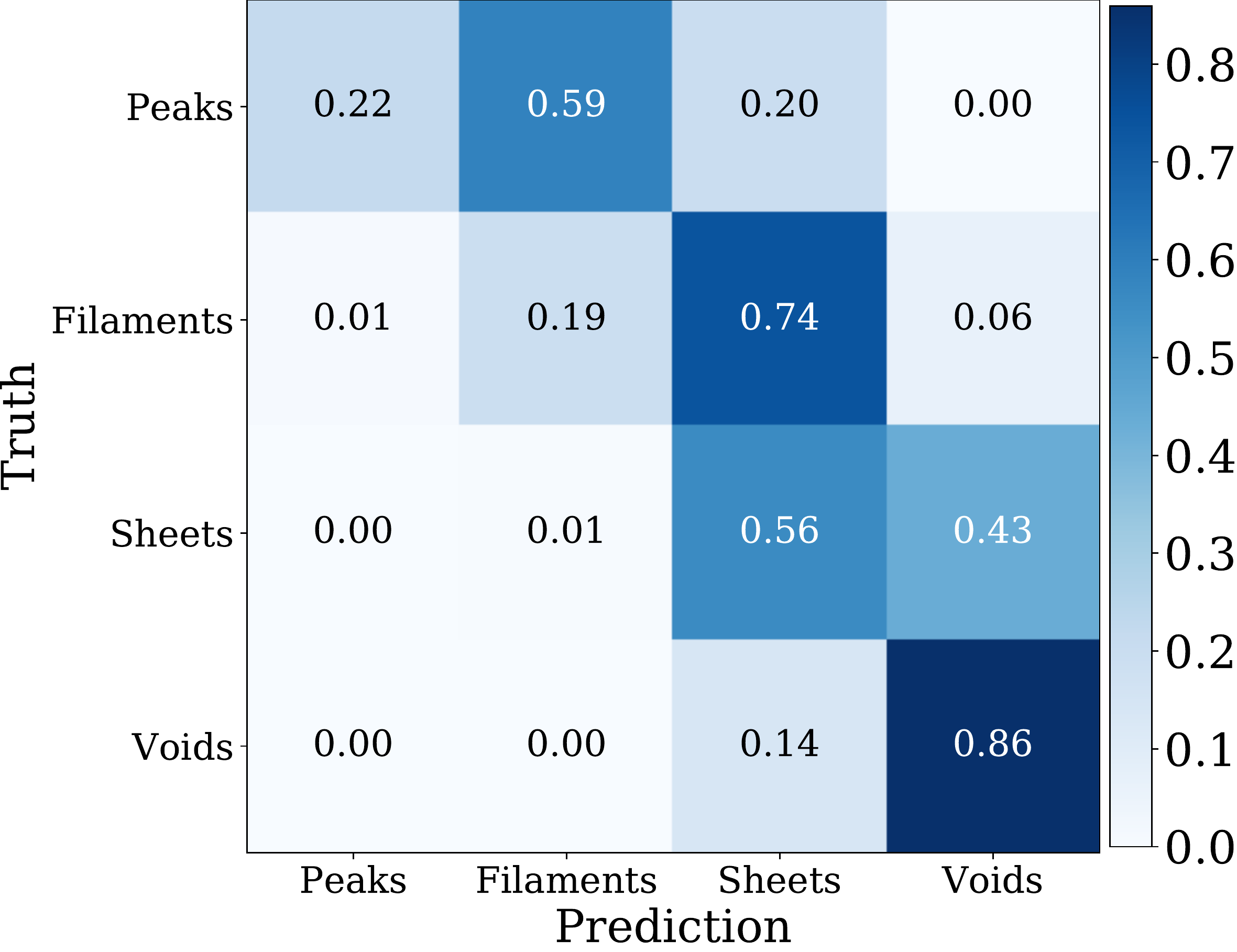}
\caption{Confusion matrix for the best Random Forest model
      \{$\lambda_{th}=0.0$, $\sigma=2.0$, $M^{*}_{\rm lim}=9$, $NE=80$, \} after training with classes downsampled to have the same number of instances as the void class.}
      \label{fig:confusion_matrix_vb}
\end{figure}

A second strategy consists in subselecting the galaxy training catalog to have roughly equal size for each environment. 
This means that we train with a number of galaxies for each class limited to have the same size as the number of void galaxies
Figure \ref{fig:confusion_matrix_vb} shows the confusion matrix for this experiment. 
The correct classification increases for voids and sheets, from $0\%$ to $86\%$ and $47\%$ to $56\%$, respectively.
However, it drops from $70\%$ to $20\%$ for peaks, and from $83\%$ to $19\%$ for filaments. 
The weighted F1 score for this case decreases from $0.73$ to $0.27$. 
We also discard this strategy.

\section{Comparison against similar approaches}\label{sec:discussion}


To elucidate the strengths of the different aspects in the approach presented here, we focus the discussion of our results on the comparison against previous work by \cite{Tsizh2019} that use a similar methodology.
Their work combines network analysis with ML to predict the environments in the cosmic web.
Their input data comes from publicly available files\footnote{\url{https://data.aip.de/projects/tracingthecosmicweb.html}} from the \emph{Tracing the Cosmic Web} comparison project \citep{Libeskind2018}.

That project performed a dark matter only simulation with $512^3$ particles within a cubic box of 290 Mpc on a side.
The halos in the simulation were identified by a Friends-Of-Friends algorithm.
Furthermore, the catalogs also include the cosmic web classification according to different algorithms.
The minimum DM halo mass included in the catalog is 10$^{11}$ M$_{\odot}$ $h^{-1}$.
This corresponds to a total of 281465 halos and a number density of $10\times 10^{-3}$ Mpc$^{-3}$, which is close to the galaxy number density in our best model with $M^{*}_{\rm lim}=9$.

\cite{Tsizh2019} built networks with fixed linking lengths in the range $l=2$-$4$ \Mpch.
From the networks they computed 10 different metrics, which, together with the mass and the
peculiar velocity, made up the features for the ML algorithms.
They also explored five different cosmic web classification algorithms.
As a score metric they use the global accuracy (the ratio of correct predictions to total number of instances).
A crucial difference with our work is that \cite{Tsizh2019} did not explore the influence of different smoothing lengths, $\sigma$, nor eigenvalue thresholds, $\lambda_{\mathrm{th}}$, in the T-Web cosmic web classifications for their dataset.
Likewise, they did not explore the influence of a dark matter halo mass cut, to produce graphs with different number density, in their prediction results.

With that setup \cite{Tsizh2019} found that the best algorithm to predict the T-Web classification was the extra gradient boosting decision trees (\texttt{xgboost}) \citep{Chen2016}.
They reported a global accuracy of 51 percent.
Unfortunately, they did not report the confusion matrix for T-Web classifications and we cannot compare the F1 scores.

If we use the same catalog they used to build the $\beta$-skeleton and then compute the features to find the best RF model we have a global accuracy of 54 percent, 3 percent points higher than \cite{Tsizh2019}.
If we also allow for cuts on halo mass in order to vary the number density we find that using only halos more massive than $10^{12}$ \Msun $h^{-1}$ increases the global accuracy to 58 percent.
Using the full setup we present here, having well resolved galaxies from the Illustris-TNG simulation, we achieve an accuracy score of 74 percent, 23 percentual points higher than \cite{Tsizh2019}.

These tests with increasing complexity tells us that there are three main factors that explain the improved performance of our approach.
First, the $\beta$-skeleton, due to its adaptive nature, seems to provide a slightly better T-Web representation than a network with a fixed linking length.
Second, the exploration in number density is fundamental to find the best match between the graph features and a given T-Web classification.
Finally, the galaxies provided by Illustris-TNG add the information needed to improve the graph match against the dark matter T-Web environment.

\section{Application to Observational Data}{\label{sec:extension}}

Applying the work presented here to observational data from large spectroscopic surveys has two main caveats to take into account: the lightcone effect and  redshift space distortions.

The lightcone effect is the time evolution of the cosmic web along the line-of-sight direction. 
For sufficiently deep surveys the cosmic web in nearby galaxies is statistically different from the cosmic web in deeper regions in the survey.
Fortunately, the cosmic web shows a slow varying evolution. 
Quantities such as the mean dark matter density and volume fraction for each one of the four cosmic web environments evolve less than $10\%$ over redshift steps of $0.2$ \citep{Cautun2014}. 
This allows for a reasonable approximation that survey slices in the radial direction with widths on the scale of $\Delta z=0.2$ should not present significant lightcone effects.

On the other hand, the radial redshift space distortions (RSD) has  two important aspects that are more difficult to overcome.
First, neighboring galaxies in redshift space may actually belong to very different environments in real space. 
In the case we use in this paper, the galaxies share the same environment on the scale of $\sigma$. This is not the case anymore once observed redshifts are translated into comoving positions.
Second, regions that are statistically isotropic, such as clusters, now show an filamentary shape along the radial direction. Furthermore, these effect has a strong dependence on the local density.

Future ML work that seeks to fully account for the lightcone and RSD effects will have to be retrained from mock catalogs that properly include these effects. 
It is also very likely that a new graph features will have to be included to account for the additional radial anisotropy.
For all these reasons, we advise that using our best model on observational data should only be considered as a zero-th order approximation to define cosmic web environments.

\section{Conclusions}\label{sec:conclusions}

Generally speaking, the methods already available in the literature to estimate the cosmic web environment of observed galaxies fall into two categories: computationally inexpensive but uncertain in their results (i.e. by directly using the galaxy number overdensity as a proxy for the dark matter overdensity); and highly accurate but computationally expensive (i.e. by performing thousands of N-body simulations to build a full dark matter reconstruction in a Bayesian framework).

In this paper we explored an approach that represents a middle ground in computational cost and accuracy in the final results.
We used the T-Web as the cosmic web definition \citep{Forero-Romero2009} and the $\beta$-skeleton graph \citep{Fang2019} to describe the relative spatial distribution of the galaxies.
The link between the T-Web and the $\beta$-graph was done through three different machine learning algorithms: Classification Trees, Extra Gradient Boosting and Random Forest.

We tested the method using data from the Illustris-TNG simulation \citep{Nelson2018,Naiman2018,Springel2018, Marinacci2018, Pillepich2018, Nelson2019} at $z=0$. 
Using the weighted F1-score as a benchmark, we first found that the Random Forest algorithm provides the best results.
Then, we showed that the most accurate predictions between the graph properties and the T-Web environment are provided for a dark matter density field smoothed over a scales $2$ Mpc and an eigenvalue threshold of $\lambda_{th}=0.0$.
The preferred eigenvalue threshold turns out to be the in ballpark range favored in previous publications for T-Web studies for the resulting classification to match the visual impression of the cosmic web \citep{Forero-Romero2009}.

For the best model the weighted F1 score is 0.728 and its global accuracy is $74\%$.
The environments ranked from higher to lower completeness are filaments ($83\%$), peaks ($70\%$), sheets ($47\%$), and voids ($0\%$).
The great difficulty to classify galaxies or isolated halos inside voids continues to be an outstanding problem as discussed in the previous publications \citep{Libeskind2018, Tsizh2019}.
In general, high completeness and F1 scores correlate with the number of instances in each class.
A larger number of instances make it easier for the algorithm to find the relevant features for its correct classification.

We compared our results against a similar methodology proposed by \cite{Tsizh2019}.
They used dark matter halos, fixed linking length network metrics and extra gradient boosting decision trees to predict the T-Web environment.
They reported a $51\%$ accuracy, meaning that our method achieves $23$ percent points more.
We showed that this improvement could be explained by the combination of three factors: the $\beta$-skeleton providing a slightly better description of the T-Web features, the meta-parameter search to find the best match between the $\beta$-skeleton and a given T-Web classification and finally, having galaxies instead of halos as tracers.

Our results provide the baseline for future work that will have to quantify the effect of redshift space distortions and survey incompleteness to predict the dark matter T-Web form survey data.
In order to facilitate future comparison and reuse of our results, we make our best-trained model public, together with a test datasets at \url{https://github.com/jsuarez314/cosmicweb_bsk}.

\section*{Data Availability}
The datasets were derived from sources in the public domain: Illustris-TNG project, \url{https://www.tng-project.org/}
and \emph{Tracing the Cosmic Web} comparison project \url{https://data.aip.de/projects/tracingthecosmicweb.htm}

The \emph{test} dataset and the best model presented in this paper are available at \url{https://github.com/jsuarez314/cosmicweb_bsk}.

\section*{Acknowledgements}

This work was supported by National SKA Program of China No. 2020SKA0110401. XDL acknowledges the support from the NSFC grant (No. 11803094), the Science and Technology Program of Guangzhou, China (No. 202002030360).

JFSP and JEFR acknowledge the support by the LACEGAL network with support from the European Union's Horizon 2020 Research and Innovation programme under the Marie Sklodowska-Curie grant agreement number 734374.

We are thankful to the community developing and maintaining open source packages fundamental to our work: numpy
\&  scipy  \citep{VanDerWalt2011},  the  Jupyter  notebook  \citep{Kluyver2016}, matplotlib \citep{Hunter2007}, corner.py \citep{Foreman-Mackey2016}, SciKit-learn \citep{Pedregosa2011} and Xgboost \citep{Chen2016}.

\bibliography{references}

\begin{thebibliography}{}
\expandafter\ifx\csname natexlab\endcsname\relax\def\natexlab#1{#1}\fi
\providecommand{\url}[1]{\href{#1}{#1}}
\providecommand{\dodoi}[1]{doi:~\href{http://doi.org/#1}{\nolinkurl{#1}}}
\providecommand{\doeprint}[1]{\href{http://ascl.net/#1}{\nolinkurl{http://ascl.net/#1}}}
\providecommand{\doarXiv}[1]{\href{https://arxiv.org/abs/#1}{\nolinkurl{https://arxiv.org/abs/#1}}}

\bibitem[{Ade {et~al.}(2016)Ade, Aghanim, Arnaud, Ashdown, Aumont, Baccigalupi,
  Banday, Barreiro, Bartlett, Bartolo, Battaner, Battye, Benabed, Beno{\^{i}}t,
  Benoit-L{\'{e}}vy, Bernard, Bersanelli, Bielewicz, Bock, Bonaldi, Bonavera,
  Bond, Borrill, Bouchet, Boulanger, Bucher, Burigana, Butler, Calabrese,
  Cardoso, Catalano, Challinor, Chamballu, Chary, Chiang, Chluba, Christensen,
  Church, Clements, Colombi, Colombo, Combet, Coulais, Crill, Curto, Cuttaia,
  Danese, Davies, Davis, {De Bernardis}, {De Rosa}, {De Zotti}, Delabrouille,
  D{\'{e}}sert, {Di Valentino}, Dickinson, Diego, Dolag, Dole, Donzelli,
  Dor{\'{e}}, Douspis, Ducout, Dunkley, Dupac, Efstathiou, Elsner, En{\ss}lin,
  Eriksen, Farhang, Fergusson, Finelli, Forni, Frailis, Fraisse, Franceschi,
  Frejsel, Galeotta, Galli, Ganga, Gauthier, Gerbino, Ghosh, Giard,
  Giraud-H{\'{e}}raud, Giusarma, Gjerl{\o}w, Gonz{\'{a}}lez-Nuevo,
  G{\'{o}}rski, Gratton, Gregorio, Gruppuso, Gudmundsson, Hamann, Hansen,
  Hanson, Harrison, Helou, Henrot-Versill{\'{e}}, Hern{\'{a}}ndez-Monteagudo,
  Herranz, Hildebrandt, Hivon, Hobson, Holmes, Hornstrup, Hovest, Huang,
  Huffenberger, Hurier, Jaffe, Jaffe, Jones, Juvela, Keih{\"{a}}nen, Keskitalo,
  Kisner, Kneissl, Knoche, Knox, Kunz, Kurki-Suonio, Lagache,
  L{\"{a}}hteenm{\"{a}}ki, Lamarre, Lasenby, Lattanzi, Lawrence, Leahy,
  Leonardi, Lesgourgues, Levrier, Lewis, Liguori, Lilje, Linden-V{\o}rnle,
  L{\'{o}}pez-Caniego, Lubin, Maci{\'{a}}s-P{\'{e}}rez, Maggio, Maino,
  Mandolesi, Mangilli, Marchini, Maris, Martin, Martinelli,
  Mart{\'{i}}nez-Gonz{\'{a}}lez, Masi, Matarrese, Mcgehee, Meinhold,
  Melchiorri, Melin, Mendes, Mennella, Migliaccio, Millea, Mitra,
  Miville-Desch{\^{e}}nes, Moneti, Montier, Morgante, Mortlock, Moss, Munshi,
  Murphy, Naselsky, Nati, Natoli, Netterfield, N{\o}rgaard-Nielsen, Noviello,
  Novikov, Novikov, Oxborrow, Paci, Pagano, Pajot, Paladini, Paoletti,
  Partridge, Pasian, Patanchon, Pearson, Perdereau, Perotto, Perrotta,
  Pettorino, Piacentini, Piat, Pierpaoli, Pietrobon, Plaszczynski,
  Pointecouteau, Polenta, Popa, Pratt, Pr{\'{e}}zeau, Prunet, Puget, Rachen,
  Reach, Rebolo, Reinecke, Remazeilles, Renault, Renzi, Ristorcelli, Rocha,
  Rosset, Rossetti, Roudier, {Rouill{\'{e}} D'orfeuil}, Rowan-Robinson,
  Rubin{\~{o}}-Mart{\'{i}}n, Rusholme, Said, Salvatelli, Salvati, Sandri,
  Santos, Savelainen, Savini, Scott, Seiffert, Serra, Shellard, Spencer,
  Spinelli, Stolyarov, Stompor, Sudiwala, Sunyaev, Sutton, Suur-Uski, Sygnet,
  Tauber, Terenzi, Toffolatti, Tomasi, Tristram, Trombetti, Tucci, Tuovinen,
  T{\"{u}}rler, Umana, Valenziano, Valiviita, {Van Tent}, Vielva, Villa, Wade,
  Wandelt, Wehus, White, White, Wilkinson, Yvon, Zacchei, \& Zonca}]{Ade2016}
Ade, P.~A., Aghanim, N., Arnaud, M., {et~al.} 2016, Astron. Astrophys., 594,
  A13, \dodoi{10.1051/0004-6361/201525830}

\bibitem[{Aikio \& Mahonen(1998)}]{Aikio1998}
Aikio, J., \& Mahonen, P. 1998, The Astrophysical Journal, 497, 534,
  \dodoi{10.1086/305509}

\bibitem[{Alam {et~al.}(2019)Alam, Zu, Peacock, \& Mandelbaum}]{Shadab2019}
Alam, S., Zu, Y., Peacock, J.~A., \& Mandelbaum, R. 2019, Mon. Not. R. Astron.
  Soc., 483, 4501, \dodoi{10.1093/mnras/sty3477}

\bibitem[{Ali {et~al.}(2012)Ali, Khan, Ahmad, \& Maqsood}]{ali2012random}
Ali, J., Khan, R., Ahmad, N., \& Maqsood, I. 2012, International Journal of
  Computer Science Issues (IJCSI), 9, 272

\bibitem[{{Alpaslan} {et~al.}(2016){Alpaslan}, {Grootes}, {Marcum}, {Popescu},
  {Tuffs}, {Bland-Hawthorn}, {Brough}, {Brown}, {Davies}, {Driver}, {Holwerda},
  {Kelvin}, {Lara-L{\'o}pez}, {L{\'o}pez-S{\'a}nchez}, {Loveday}, {Moffett},
  {Taylor}, {Owers}, \& {Robotham}}]{Alpaslan2016}
{Alpaslan}, M., {Grootes}, M., {Marcum}, P.~M., {et~al.} 2016, \mnras, 457,
  2287, \dodoi{10.1093/mnras/stw134}

\bibitem[{Arag{\'{o}}n-Calvo {et~al.}(2010)Arag{\'{o}}n-Calvo, Platen, {Van De
  Weygaert}, \& Szalay}]{Aragon-Calvo2010}
Arag{\'{o}}n-Calvo, M.~A., Platen, E., {Van De Weygaert}, R., \& Szalay, A.~S.
  2010, Astrophysical Journal, 723, 364, \dodoi{10.1088/0004-637X/723/1/364}

\bibitem[{{Bloom} {et~al.}(2012){Bloom}, {Richards}, {Nugent}, {Quimby},
  {Kasliwal}, {Starr}, {Poznanski}, {Ofek}, {Cenko}, {Butler}, {Kulkarni},
  {Gal-Yam}, \& {Law}}]{Bloom2012}
{Bloom}, J.~S., {Richards}, J.~W., {Nugent}, P.~E., {et~al.} 2012, \pasp, 124,
  1175, \dodoi{10.1086/668468}

\bibitem[{Bond {et~al.}(1996)Bond, Kofman, \& Pogosyan}]{Bond1996}
Bond, J.~R., Kofman, L., \& Pogosyan, D. 1996, Nature, 380, 603,
  \dodoi{10.1038/380603a0}

\bibitem[{Bonnaire {et~al.}(2020)Bonnaire, Aghanim, Decelle, \&
  Douspis}]{Bonnaire2020}
Bonnaire, T., Aghanim, N., Decelle, A., \& Douspis, M. 2020, Astronomy and
  Astrophysics, 637, \dodoi{10.1051/0004-6361/201936859}

\bibitem[{Bos {et~al.}(2014)Bos, {Van De Weygaert}, Kitaura, \&
  Cautun}]{Bos2014}
Bos, E.~G., {Van De Weygaert}, R., Kitaura, F., \& Cautun, M. 2014, in Proc.
  Int. Astron. Union, Vol.~11 (Cambridge University Press), 271--288

\bibitem[{{Buncher} \& {Carrasco Kind}(2020)}]{2020MNRAS.497.5041B}
{Buncher}, B., \& {Carrasco Kind}, M. 2020, \mnras, 497, 5041,
  \dodoi{10.1093/mnras/staa2008}

\bibitem[{Burchett {et~al.}(2020)Burchett, Elek, Tejos, Prochaska, Tripp,
  Bordoloi, \& Forbes}]{Burchett2020}
Burchett, J.~N., Elek, O., Tejos, N., {et~al.} 2020, Astrophys. J., 891, L35,
  \dodoi{10.3847/2041-8213/ab700c}

\bibitem[{Bustamante \& Forero-Romero(2015)}]{Bustamante2015}
Bustamante, S., \& Forero-Romero, J.~E. 2015, Mon. Not. R. Astron. Soc., 453,
  497, \dodoi{10.1093/mnras/stv1637}

\bibitem[{Cautun {et~al.}(2013)Cautun, van~de Weygaert, \& Jones}]{Cautun2013}
Cautun, M., van~de Weygaert, R., \& Jones, B.~J. 2013, Monthly Notices of the
  Royal Astronomical Society, 429, 1286, \dodoi{10.1093/mnras/sts416}

\bibitem[{Cautun {et~al.}(2014)Cautun, {Van De Weygaert}, Jones, \&
  Frenk}]{Cautun2014}
Cautun, M., {Van De Weygaert}, R., Jones, B.~J., \& Frenk, C.~S. 2014, Mon.
  Not. R. Astron. Soc., 441, 2923, \dodoi{10.1093/mnras/stu768}

\bibitem[{Chen \& Guestrin(2016)}]{Chen2016}
Chen, T., \& Guestrin, C. 2016, Proceedings of the ACM SIGKDD International
  Conference on Knowledge Discovery and Data Mining, 13-17-August-2016, 785,
  \dodoi{10.1145/2939672.2939785}

\bibitem[{Chen {et~al.}(2015)Chen, Ho, Freeman, Genovese, \&
  Wasserman}]{Chen2015}
Chen, Y.~C., Ho, S., Freeman, P.~E., Genovese, C.~R., \& Wasserman, L. 2015,
  Mon. Not. R. Astron. Soc., 454, 1140, \dodoi{10.1093/mnras/stv1996}

\bibitem[{Eardley {et~al.}(2015)Eardley, Peacock, McNaught-Roberts, Heymans,
  Norberg, Alpaslan, Baldry, Bland-Hawthorn, Brough, Cluver, Driver, Farrow,
  Liske, Loveday, \& Robotham}]{Eardley2015}
Eardley, E., Peacock, J.~A., McNaught-Roberts, T., {et~al.} 2015, Mon. Not. R.
  Astron. Soc., 448, 3665, \dodoi{10.1093/mnras/stv237}

\bibitem[{Elyiv {et~al.}(2015)Elyiv, Marulli, Pollina, Baldi, Branchini,
  Cimatti, \& Moscardini}]{Elyiv2015}
Elyiv, A., Marulli, F., Pollina, G., {et~al.} 2015, Monthly Notices of the
  Royal Astronomical Society, 448, 642, \dodoi{10.1093/mnras/stv043}

\bibitem[{Fang {et~al.}(2019)Fang, Forero-Romero, Rossi, Li, \&
  Feng}]{Fang2019}
Fang, F., Forero-Romero, J., Rossi, G., Li, X.~D., \& Feng, L.~L. 2019, Mon.
  Not. R. Astron. Soc., 485, 5276, \dodoi{10.1093/mnras/stz773}

\bibitem[{Foreman-Mackey(2016)}]{Foreman-Mackey2016}
Foreman-Mackey, D. 2016, J. Open Source Softw., 1, 24,
  \dodoi{10.21105/joss.00024}

\bibitem[{Forero-Romero {et~al.}(2009)Forero-Romero, Hoffman, Gottl{\"{o}}ber,
  Klypin, \& Yepes}]{Forero-Romero2009}
Forero-Romero, J.~E., Hoffman, Y., Gottl{\"{o}}ber, S., Klypin, A., \& Yepes,
  G. 2009, Mon. Not. R. Astron. Soc., 396, 1815,
  \dodoi{10.1111/j.1365-2966.2009.14885.x}

\bibitem[{Garcia-Alvarado {et~al.}(2020)Garcia-Alvarado, Li, \&
  Forero-Romero}]{Garcia-Alvarado2020}
Garcia-Alvarado, M.~V., Li, X.~D., \& Forero-Romero, J.~E. 2020, Monthly
  Notices of the Royal Astronomical Society: Letters, 498, L145,
  \dodoi{10.1093/mnrasl/slaa145}

\bibitem[{Hahn {et~al.}(2007)Hahn, Porciani, Carollo, \& Dekel}]{Hahn2007}
Hahn, O., Porciani, C., Carollo, C.~M., \& Dekel, A. 2007, Mon. Not. R. Astron.
  Soc., 375, 489, \dodoi{10.1111/j.1365-2966.2006.11318.x}

\bibitem[{Horowitz {et~al.}(2019)Horowitz, Lee, White, Krolewski, \&
  Ata}]{Horowitz2019}
Horowitz, B., Lee, K.-G., White, M., Krolewski, A., \& Ata, M. 2019, Astrophys.
  J., 887, 61, \dodoi{10.3847/1538-4357/ab4d4c}

\bibitem[{Hui {et~al.}(2018)Hui, Aragon, Cui, \& Flegal}]{Hui2018}
Hui, J., Aragon, M., Cui, X., \& Flegal, J.~M. 2018, Mon. Not. R. Astron. Soc.,
  475, 4494, \dodoi{10.1093/mnras/stx3235}

\bibitem[{Hunter(2007)}]{Hunter2007}
Hunter, J.~D. 2007, Comput. Sci. Eng., 9, 99, \dodoi{10.1109/MCSE.2007.55}

\bibitem[{Jasche {et~al.}(2010)Jasche, Kitaura, Li, \& En{\ss}lin}]{Jasche2010}
Jasche, J., Kitaura, F.~S., Li, C., \& En{\ss}lin, T.~A. 2010, Mon. Not. R.
  Astron. Soc., 409, 355, \dodoi{10.1111/j.1365-2966.2010.17313.x}

\bibitem[{Jasche \& Wandelt(2013)}]{Jasche2013a}
Jasche, J., \& Wandelt, B.~D. 2013, Mon. Not. R. Astron. Soc., 432, 894,
  \dodoi{10.1093/mnras/stt449}

\bibitem[{Kirkpatrick \& Radke(1985)}]{Kirkpatrick1985}
Kirkpatrick, D.~G., \& Radke, J.~D. 1985, Mach. Intell. Pattern Recognit., 2,
  217, \dodoi{10.1016/B978-0-444-87806-9.50013-X}

\bibitem[{Kluyver {et~al.}(2016)Kluyver, Ragan-kelley, P{\'{e}}rez, Granger,
  Bussonnier, Frederic, Kelley, Hamrick, Grout, Corlay, Ivanov, Avila, Abdalla,
  Willing, \& {Development Team}}]{Kluyver2016}
Kluyver, T., Ragan-kelley, B., P{\'{e}}rez, F., {et~al.} 2016, in Position.
  Power Acad. Publ. Play. Agents Agendas, ed. F.~Loizides \& B.~Schmidt (IOS
  Press), 87--90

\bibitem[{Leclercq {et~al.}(2015)Leclercq, Jasche, \&
  Wandelt}]{LeclercqJasche2015}
Leclercq, F., Jasche, J., \& Wandelt, B. 2015, Astron. Astrophys., 576,
  \dodoi{10.1051/0004-6361/201526006}

\bibitem[{Libeskind {et~al.}(2018)Libeskind, van~de Weygaert, Cautun, Falck,
  Tempel, Abel, Alpaslan, Arag{\'{o}}n-Calvo, Forero-Romero, Gonzalez,
  Gottl{\"{o}}ber, Hahn, Hellwing, Hoffman, Jones, Kitaura, Knebe, Manti,
  Neyrinck, Nuza, Padilla, Platen, Ramachandra, Robotham, Saar, Shandarin,
  Steinmetz, Stoica, Sousbie, \& Yepes}]{Libeskind2018}
Libeskind, N.~I., van~de Weygaert, R., Cautun, M., {et~al.} 2018, Mon. Not. R.
  Astron. Soc., 473, 1195, \dodoi{10.1093/mnras/stx1976}

\bibitem[{Luber {et~al.}(2019)Luber, van Gorkom, Hess, Pisano, Fern{\'{a}}ndez,
  \& Momjian}]{Luber2019}
Luber, N., van Gorkom, J.~H., Hess, K.~M., {et~al.} 2019, Astron. J., 157, 254,
  \dodoi{10.3847/1538-3881/ab1b6e}

\bibitem[{Marinacci {et~al.}(2018)Marinacci, Vogelsberger, Pakmor, Torrey,
  Springel, Hernquist, Nelson, Weinberger, Pillepich, Naiman, \&
  Genel}]{Marinacci2018}
Marinacci, F., Vogelsberger, M., Pakmor, R., {et~al.} 2018, Monthly Notices of
  the Royal Astronomical Society, 480, 5113, \dodoi{10.1093/mnras/sty2206}

\bibitem[{Mu{\~{n}}oz-Cuartas {et~al.}(2011)Mu{\~{n}}oz-Cuartas, M{\"{u}}ller,
  \& Forero-Romero}]{Munoz-Cuartas2011}
Mu{\~{n}}oz-Cuartas, J.~C., M{\"{u}}ller, V., \& Forero-Romero, J.~E. 2011,
  Mon. Not. R. Astron. Soc., 417, 1303,
  \dodoi{10.1111/j.1365-2966.2011.19344.x}

\bibitem[{Naiman {et~al.}(2018)Naiman, Pillepich, Springel, Ramirez-Ruiz,
  Torrey, Vogelsberger, Pakmor, Nelson, Marinacci, Hernquist, Weinberger, \&
  Genel}]{Naiman2018}
Naiman, J.~P., Pillepich, A., Springel, V., {et~al.} 2018, Monthly Notices of
  the Royal Astronomical Society, 477, 1206, \dodoi{10.1093/mnras/sty618}

\bibitem[{{Neira} {et~al.}(2020){Neira}, {G{\'o}mez}, {Su{\'a}rez-P{\'e}rez},
  {G{\'o}mez}, {Reyes}, {Hoyos}, {Arbel{\'a}ez}, \&
  {Forero-Romero}}]{Neira2020}
{Neira}, M., {G{\'o}mez}, C., {Su{\'a}rez-P{\'e}rez}, J.~F., {et~al.} 2020,
  \apjs, 250, 11, \dodoi{10.3847/1538-4365/aba267}

\bibitem[{Nelson {et~al.}(2018)Nelson, Pillepich, Springel, Weinberger,
  Hernquist, Pakmor, Genel, Torrey, Vogelsberger, Kauffmann, Marinacci, \&
  Naiman}]{Nelson2018}
Nelson, D., Pillepich, A., Springel, V., {et~al.} 2018, Monthly Notices of the
  Royal Astronomical Society, 475, 624, \dodoi{10.1093/mnras/stx3040}

\bibitem[{Nelson {et~al.}(2019)Nelson, Springel, Pillepich, Rodriguez-Gomez,
  Torrey, Genel, Vogelsberger, Pakmor, Marinacci, Weinberger, Kelley, Lovell,
  Diemer, \& Hernquist}]{Nelson2019}
Nelson, D., Springel, V., Pillepich, A., {et~al.} 2019, Comput. Astrophys.
  Cosmol., 6, \dodoi{10.1186/s40668-019-0028-x}

\bibitem[{Neyrinck(2008)}]{Neyrinck2008}
Neyrinck, M.~C. 2008, Mon. Not. R. Astron. Soc., 386, 2101,
  \dodoi{10.1111/j.1365-2966.2008.13180.x}

\bibitem[{Nordhausen(2009)}]{Nordhausen2009}
Nordhausen, K. 2009, International Statistical Review, 77, 482,
  \dodoi{10.1111/j.1751-5823.2009.00095_18.x}

\bibitem[{Novikov {et~al.}(2006)Novikov, Colombi, \& Dor{\'{e}}}]{Novikov2003}
Novikov, D., Colombi, S., \& Dor{\'{e}}, O. 2006, Mon. Not. R. Astron. Soc.,
  366, 1201, \dodoi{10.1111/j.1365-2966.2005.09925.x}

\bibitem[{Padilla {et~al.}(2005)Padilla, Ceccarelli, \& Lambas}]{Padilla2005}
Padilla, N.~D., Ceccarelli, L., \& Lambas, D.~G. 2005, Monthly Notices of the
  Royal Astronomical Society, 363, 977,
  \dodoi{10.1111/j.1365-2966.2005.09500.x}

\bibitem[{Pedregosa {et~al.}(2011)Pedregosa, Varoquaux, Gramfort, Michel,
  Thirion, Grisel, Blondel, Prettenhofer, Weiss, Dubourg, Vanderplas, Passos,
  Cournapeau, Brucher, Perrot, \& Duchesnay}]{Pedregosa2011}
Pedregosa, F., Varoquaux, G., Gramfort, A., {et~al.} 2011, J. Mach. Learn.
  Res., 12, 2825.
\newblock \doarXiv{1201.0490}

\bibitem[{Pillepich {et~al.}(2018{\natexlab{a}})Pillepich, Nelson, Hernquist,
  Springe, {R{\"{u}}diger Pakmor}, Torrey, Weinberger, Gene, Naiman, Marinacci,
  \& Vogelsberger}]{Pillepich2018}
Pillepich, A., Nelson, D., Hernquist, L., {et~al.} 2018{\natexlab{a}}, Monthly
  Notices of the Royal Astronomical Society, 475, 648,
  \dodoi{10.1093/mnras/stx3112}

\bibitem[{Pillepich {et~al.}(2018{\natexlab{b}})Pillepich, Springel, Nelson,
  Genel, Naiman, Pakmor, Hernquist, Torrey, Vogelsberger, Weinberger, \&
  Marinacci}]{Pillepich2018a}
Pillepich, A., Springel, V., Nelson, D., {et~al.} 2018{\natexlab{b}}, Monthly
  Notices of the Royal Astronomical Society, 473, 4077,
  \dodoi{10.1093/mnras/stx2656}

\bibitem[{Platen {et~al.}(2007)Platen, {Van De Weygaert}, \&
  Jones}]{Platen2007}
Platen, E., {Van De Weygaert}, R., \& Jones, B.~J. 2007, Mon. Not. R. Astron.
  Soc., 380, 551, \dodoi{10.1111/j.1365-2966.2007.12125.x}

\bibitem[{Schmalzing {et~al.}(1999)Schmalzing, Buchert, Melott, Sahni,
  Sathyaprakash, \& Shandarin}]{Schmalzing1999}
Schmalzing, J., Buchert, T., Melott, A.~L., {et~al.} 1999, Astrophys. J., 526,
  568, \dodoi{10.1086/308039}

\bibitem[{Sousbie(2011)}]{Sousbie2010}
Sousbie, T. 2011, Mon. Not. R. Astron. Soc., 414, 350,
  \dodoi{10.1111/j.1365-2966.2011.18394.x}

\bibitem[{Springel(2011)}]{Springel2011}
Springel, V. 2011, Proc. Int. Astron. Union, 6, 203,
  \dodoi{10.1017/S1743921311000378}

\bibitem[{Springel {et~al.}(2018)Springel, Pakmor, Pillepich, Weinberger,
  Nelson, Hernquist, Vogelsberger, Genel, Torrey, Marinacci, \&
  Naiman}]{Springel2018}
Springel, V., Pakmor, R., Pillepich, A., {et~al.} 2018, Monthly Notices of the
  Royal Astronomical Society, 475, 676, \dodoi{10.1093/mnras/stx3304}

\bibitem[{Stoica {et~al.}(2007)Stoica, Mart{\'{i}}nez, \& Saar}]{Stoica2007}
Stoica, R.~S., Mart{\'{i}}nez, V.~J., \& Saar, E. 2007, Journal of the Royal
  Statistical Society. Series C: Applied Statistics, 56, 459,
  \dodoi{10.1111/j.1467-9876.2007.00587.x}

\bibitem[{Sutter {et~al.}(2015)Sutter, Lavaux, Hamaus, Pisani, Wandelt, Warren,
  Villaescusa-Navarro, Zivick, Mao, \& Thompson}]{Sutter2015}
Sutter, P.~M., Lavaux, G., Hamaus, N., {et~al.} 2015, Astronomy and Computing,
  9, 1, \dodoi{10.1016/j.ascom.2014.10.002}

\bibitem[{Tojeiro {et~al.}(2017)Tojeiro, Eardley, Peacock, Norberg, Alpaslan,
  Driver, Henriques, Hopkins, Kafle, Robotham, Thomas, Tonini, \&
  Wild}]{Tojeiro2017}
Tojeiro, R., Eardley, E., Peacock, J.~A., {et~al.} 2017, Mon. Not. R. Astron.
  Soc., 470, 3720, \dodoi{10.1093/mnras/stx1466}

\bibitem[{Tsizh {et~al.}(2020)Tsizh, Novosyadlyj, Holovatch, \&
  Libeskind}]{Tsizh2019}
Tsizh, M., Novosyadlyj, B., Holovatch, Y., \& Libeskind, N.~I. 2020, Mon. Not.
  R. Astron. Soc., 495, 1311, \dodoi{10.1093/mnras/staa1030}

\bibitem[{{Van Der Walt} {et~al.}(2011){Van Der Walt}, Colbert, \&
  Varoquaux}]{VanDerWalt2011}
{Van Der Walt}, S., Colbert, S.~C., \& Varoquaux, G. 2011, Comput. Sci. Eng.,
  13, 22, \dodoi{10.1109/MCSE.2011.37}

\bibitem[{Vogelsberger {et~al.}(2014)Vogelsberger, Genel, Springel, Torrey,
  Sijacki, Xu, Snyder, Bird, Nelson, \& Hernquist}]{Vogelsberger2014}
Vogelsberger, M., Genel, S., Springel, V., {et~al.} 2014, Nature, 509, 177,
  \dodoi{10.1038/nature13316}

\bibitem[{Wang {et~al.}(2009)Wang, Mo, Jing, Guo, {Van Den Bosch}, \&
  Yang}]{Wang2009}
Wang, H., Mo, H.~J., Jing, Y.~P., {et~al.} 2009, Mon. Not. R. Astron. Soc.,
  394, 398, \dodoi{10.1111/j.1365-2966.2008.14301.x}

\bibitem[{Weinberger {et~al.}(2017)Weinberger, Springel, Hernquist, Pillepich,
  Marinacci, Pakmor, Nelson, Genel, Vogelsberger, Naiman, \&
  Torrey}]{Weinberger2017}
Weinberger, R., Springel, V., Hernquist, L., {et~al.} 2017, Monthly Notices of
  the Royal Astronomical Society, 465, 3291, \dodoi{10.1093/mnras/stw2944}

\bibitem[{White {et~al.}(1987)White, Frenk, Davis, \& Efstathiou}]{White1987}
White, S. D.~M., Frenk, C.~S., Davis, M., \& Efstathiou, G. 1987, Astrophys.
  J., 313, 505, \dodoi{10.1086/164990}

\bibitem[{Xu {et~al.}(2019)Xu, Cisewski-Kehe, Green, \& Nagai}]{Xu2019}
Xu, X., Cisewski-Kehe, J., Green, S.~B., \& Nagai, D. 2019, Astronomy and
  Computing, 27, 34, \dodoi{10.1016/j.ascom.2019.02.003}

\bibitem[{Zel'Dovich {et~al.}(1970)Zel'Dovich, Shandarin, \&
  Sunyaev}]{ZelDovich1970}
Zel'Dovich, Y., Shandarin, S., \& Sunyaev, R. 1970, Astron. Astrophys., 500, 13

\bibitem[{Zhang {et~al.}(2009)Zhang, Yang, Faltenbacher, Springel, Lin, \&
  Wang}]{Zhang2009}
Zhang, Y., Yang, X., Faltenbacher, A., {et~al.} 2009, Astrophys. J., 706, 747,
  \dodoi{10.1088/0004-637X/706/1/747}

\end{thebibliography}
\bibliographystyle{aasjournal}

\end{document}